\documentclass[preprint,12pt]{elsarticle}

\usepackage{amssymb}
\usepackage{graphicx}
\usepackage{tabularx}
\usepackage{booktabs}
\usepackage{pstricks,pst-node,pst-text,pst-3d}
\usepackage{psfrag}

\journal{Nuclear Physics B}

\begin{document}

\begin{frontmatter}



\title{A model of the reflection distribution in the vacuum ultra violet region}


\author[label1,label2]{C. Silva}
\ead{claudio@lipc.fis.uc.pt}
\author[label1,label2]{J. Pinto da Cunha}
\author[label1]{A. Pereira}
\author[label1,label2]{M. I. Lopes}
\author[label1,label2]{V. Chepel}
\author[label1,label2]{V. Solovov}
\author[label1,label2]{F. Neves}

\address[label1]{LIP-Coimbra
\\ Department of Physics, University of Coimbra
\\ P-3004 516 Coimbra, Portugal}
\address[label2]{Department of Physics, University of Coimbra
\\ P-3004 516 Coimbra, Portugal}

\begin{abstract}
A reflection model with three components, a specular spike,
a specular lobe and a diffuse lobe is discussed. This model was
successfully applied to describe reflection of 
xenon scintillation light ($\lambda=175\,$nm)
by PTFE and other fluoropolymers
and can be used for Monte Carlo simulation and analysis of scintillation
detectors. The measured data favors a Trowbridge-Reitz distribution
function of ellipsoidal micro-surfaces.
The intensity of the coherent reflection increases with increasing angle of incidence,
as expected, since the surface appears smoother at grazing
angles. The total reflectance obtained for PTFE is about
70\% for VUV light at normal incidence in vacuum and estimated to be
up to
100\% in contact with liquid xenon.

\end{abstract}


\end{frontmatter}

\section{Introduction}

The scintillation light emitted in vacuum ultra-violet (VUV) region
by noble gases/liquids (namely xenon, neon and
argon) is used in particle physics detectors to detect very small energy depositions \cite{scin}.
It is therefore desirable to know in detail the VUV reflectance inside these systems.
Moreover, this knowledge is also relevant to applications of VUV sources in
synchrotron ultraviolet radiation facilities.

Owing to a very low outgassing, chemical stability and high
reflectance, PTFE (polytetrafluoroethylene) is a common choice for 
the inner walls of VUV systems \cite{zepii},\cite{EXO}. However, PTFE 
is an inhomogeneous dielectric whose reflection pattern is not
sufficiently known in the ultraviolet.
In this work we study the reflectance distributions of some
fluoropolymers, namely PTFE, ETFE, FEP and PFA.
Most of the existing reflection distribution studies 
have been developed in the framework of computer graphics and
 computer vision and tested only for visible light \cite{oren}. 
 Experimental reflection data in the VUV region is 
 difficult to obtain due to the light absorption by the oxygen molecules in air.
We measured the reflectance of these fluoropolymers with 
xenon scintillation light (175 nm), using a specially built
scatterometer, in which the atmospheric air can be replaced by a
  non absorbing gas such as pure nitrogen or argon \cite{silva}.
Here we show data on the reflection by these surfaces along with 
the results of a model that describes effectively the observations.


The roughness of the surface, which has effect on both diffuse and specular reflections, 
is described as an ensemble of micro-surfaces randomly oriented in space, following a 
certain probability distribution function related to the surface
structure \cite{sparrow}. 
A geometrical attenuation factor accounts for shadowing and masking by different micro-surfaces, 
especially at low grazing angles.


We conclude that three parameters suffice to describe the observed reflectance of unpolished
samples of the aforementioned fluoropolymers, namely: 
i) the index of refraction, ii) the albedo of the surface and iii) a
roughness parameter \cite{silva2}. 
This model has been validated with data of various wavelengths, including VUV light from the 
scintillation of xenon (175 nm) and light from UV and green LEDs. 
The reflectance distributions of the samples were measured
both in and out of the plane of incidence.
It was found that smoother surfaces are better described if a specular reflection 
spike is also included to account for coherent reflection at the surface average plane.

Here we present a reflectance model of rough surfaces, considering
that the light reflection is both
diffuse and specular, with three components: i) the albedo due to
internal sub-surface scattering of light, ii) the reflection at the surface by a myriad 
of random faces and iii) the coherent reflection spike.

We conclude that although the reflectance of the PTFE is mainly diffuse, 
the intensity of both the specular lobe and the coherent spike
can be significant in the VUV, specially  
at grazing angles. 
Hence, this effect should be taken into account in detailed analyses of 
the light collection in scintillation detectors and other systems.

This work aimed primarily at describing the reflectance distributions of dielectric surfaces, 
namely in the VUV region, in a way suitable to be used in simulations
and analysis of the 
light propagation through the volume of a scintillation chamber, 
up to the point of detection or absorption. 
In view of these applications this model was added as a class of the 
Geant4 Monte Carlo simulation toolkit \cite{geant4} with the effect
of improving the quality of the the simulations of experiments
that rely on detection of the VUV light.


\section{Modelling the Reflectance}
\begin{figure}[!t]
\centering

\includegraphics[width= 0.85\linewidth]{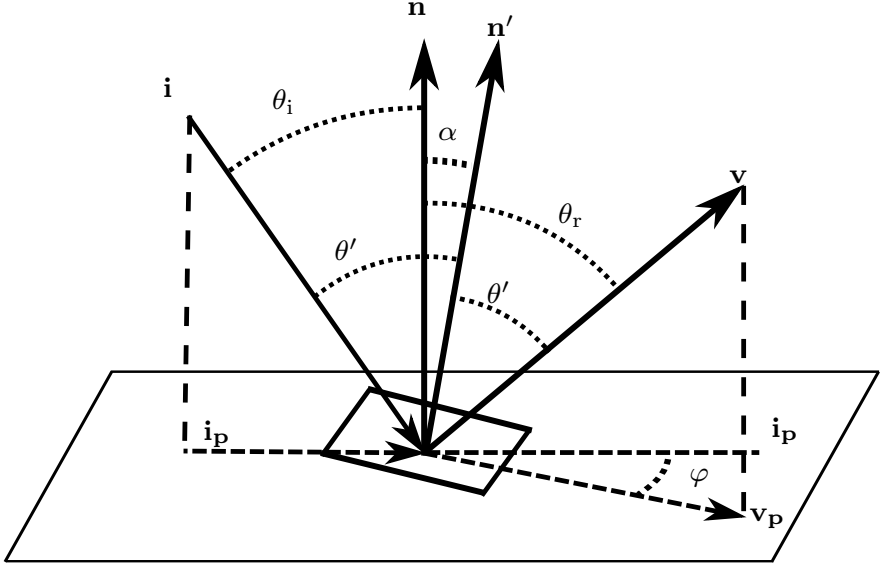}
\caption{The system of coordinates and directions.
 The vector ${\bf i}$ represents
  the direction of incident photons, ${\bf v}$ is the viewing
  direction. Primed angles are measured relatively to the normal to
  the local micro-surface, $\mathbf{n}'$, at the scattering point.}

\label{sy_model}
\end{figure}

The reflection pattern produced by a surface can be studied using the
bidirectional reflectance intensity distribution function (BRIDF).
This function
is defined as the ratio between the intensity reflected along a viewing direction, {\bf v},
and the radiation flux, $\Phi_{i}$, incident at the surface along the direction {\bf i} \cite{nicodemus}
\begin{equation}
\varrho= \frac{1}{\Phi_{i}}
\frac{\mathrm{d}\Phi_{r}}{\mathrm{d}\Omega_{r}}\qquad [1/\mathrm{sr}]
\end{equation}
where $\mathrm{d}\Phi_{r}/\mathrm{d}\Omega$ is the flux reflected per 
solid angle towards the viewing direction (see figure \ref{sy_model}).

The reflectance distribution was modeled by a function $\varrho$ comprising 
three different contributions: the
diffuse lobe $\varrho_{D}$, the specular lobe $\varrho_{S}$ and the coherent specular
spike $\varrho_{C}$,
\begin{equation}
\varrho=\varrho_{D}+\varrho_{S}+\varrho_{C}
\end{equation}
In a dielectric the diffuse lobe is associated to internal
scattering of the light that penetrates into the
material and is scattered by sub-surface inhomogeneities back into
the incoming medium. For smooth surfaces this process is approximately lambertian,
corrected by the Fresnel equations of reflection \cite{wolff},
\begin{equation}
\varrho_{D}=\frac{\rho_{L}}{\pi}\cos{\theta_{r}}\; \left\{1-F\left(\theta_{i},n/n_{0}\right)\right\}
\; \left\{1-F\left(\theta_{r},n_{0}/n\right)\right\}\nonumber
\end{equation}
where $F$ is Fresnel reflection coefficient
, $n_0$ and $n$ are the indices of refraction above an below the surface and 
$\rho_{L}$ is the albedo of the surface, i.e.
the fraction of the light which is not absorbed during the internal scattering. 

The surface roughness, though small, is important to the specular reflected light,
including the coherently reflected spike.
To describe the lobe of specular reflection we adopt a geometric reflection approach in which
specular reflection occurs at a myriad of micro-surfaces randomly oriented in space, with
some probability distribution function, $P$ \cite{sparrow}.
The corresponding BRIDF function can be put in the form \cite{silva2},
\begin{equation}
\varrho_S=\frac{\left(1-C\right)FPG}{4\cos{\theta_{i}}}
\end{equation}
where $G$ is a geometrical factor to account for the
effects of shadowing between micro-surfaces and $C$ is a weighting
factor corresponding to the fraction 
of light undergoing coherent reflection.
The factor $1/4\cos\theta_i$ relates the geometry of the solid angle made by
the normals of micro-surfaces with the solid angle to where they
reflect \cite{silva2}\cite{walter}.
The above process is in competition with the coherent reflection
responsible for a specular spike in the
reflectance distribution of the surface. This is a pure wave
phenomenon and as such
cannot be predicted by a geometrical approach to the reflection \cite{ogilvy}.
The coherent contribution can be written as
\begin{equation}
\varrho_C=CFG
\end{equation}
where $C$ measures the importance of the coherent reflection. 

 The coherent reflection wave can be predicted by the model of
Beckmann-Spizzichino.
However, this calculation uses conductor boundary conditions at the surface, 
which may not be satisfactory on a dielectric interface. 
 Notwithstanding, this model predicts that the intensity of the
specular spike behave as an exponential function of
the surface roughness and of the incident and reflected angles \cite{nayer}, as 
\begin{equation}
C=\exp{\left\{-\left[\frac{4\pi}{\lambda}\sigma_{h}\left(\cos{\theta_{i}}+\cos{\theta_{r}}\right)\right]^{2}\right\}}
\end{equation}
where $\theta_i$ and $\theta_r$ are the incident and reflection angles respectively, 
$\sigma_h$ is the standard deviation of the distribution of the height irregularities of the
surface, assumed Gaussian, and $\lambda$ is the wavelength of the
radiation. 
Thus, at large angles of incidence the surface
appears smoother and the importance of the specular spike
increases. However, the fits to the reflectance of the measured
fluoropolymer dielectrics clearly suggest the following dependency
\begin{equation}
C=\exp{\left\{-\frac{K}{2}\left(\cos\theta_i+\cos\theta_{r}\right)\right\}}
\end{equation}
where $K$ is a constant to be fitted to the data.


As stated above, the shape of the specular lobe is related with the surface structure function.
The surface is modeled by an ensemble of micro-surfaces oriented at
random whose normals make angles $\alpha$
with the macroscopic normal and are distributed according to some
(usually unknown) probability distribution function $P(\alpha)$ (see figure \ref{sy_model}).
We compared the predictions based on  two possible probability distribution functions: i) the
Cook-Torrance function for V shaped micro-surfaces \cite{cook},
$P_{CT}$, which is approximately gaussian for small roughness, and 
ii) the function deduced by Trowbridge-Reitz for 
ellipsoidal micro-surfaces \cite{reitz}, $P_{TR}$:
\begin{eqnarray}
P_{TR}\left(\alpha;\gamma\right) &=&
\frac{\gamma^{2}}{\pi\cos^{4}\alpha\left(\gamma^{2}+\tan^{2}{\alpha}\right)^{2}} \label{beck}\\
P_{CT}\left(\alpha;m\right) &=& \frac{1}{\pi m^{2}\cos^{4}\alpha}
\exp{\left(-\tan^{2}\alpha/m\right)}\label{reitz}
\end{eqnarray}
The two functions are normalized to unity and the parameters $\gamma$ and $m$ measure, in each case, the 
roughness of the surface.
 The fits to the measured
data suggest that the Trowbridge-Reitz function is more suitable to
represent the probability distribution of the micro-surfaces (see figure \ref{normal_vstr}). Therefore,
in what follows  we adopt this probability function unless otherwise
stated. 



The geometrical factor $G$ mentioned above, that corrects for
shadowing and masking effects by adjacent micro-facets, was borrowed from
Smith formulae \cite{smith}, 
$$G(\theta_i,\theta_r,\varphi_r)\simeq 
H(\theta'_i-\pi/2)H(\theta'_r-\pi/2)G'(\theta_i)G'(\theta_r)\nonumber$$
where $H$ is the Heaviside step function and $G'$ is a function which
is given by
\begin{equation}
G\left(\theta\right) = \frac{2}{1+\sqrt{1+\gamma^{2}\tan^{2}\theta}}
\end{equation}
if the surface irregularities are distributed according to the
Trowbridge-Reitz probability function \cite{walter}.

\section{Experimental results}

\begin{figure*}[!t]
\centering
\includegraphics[width=6.0in]{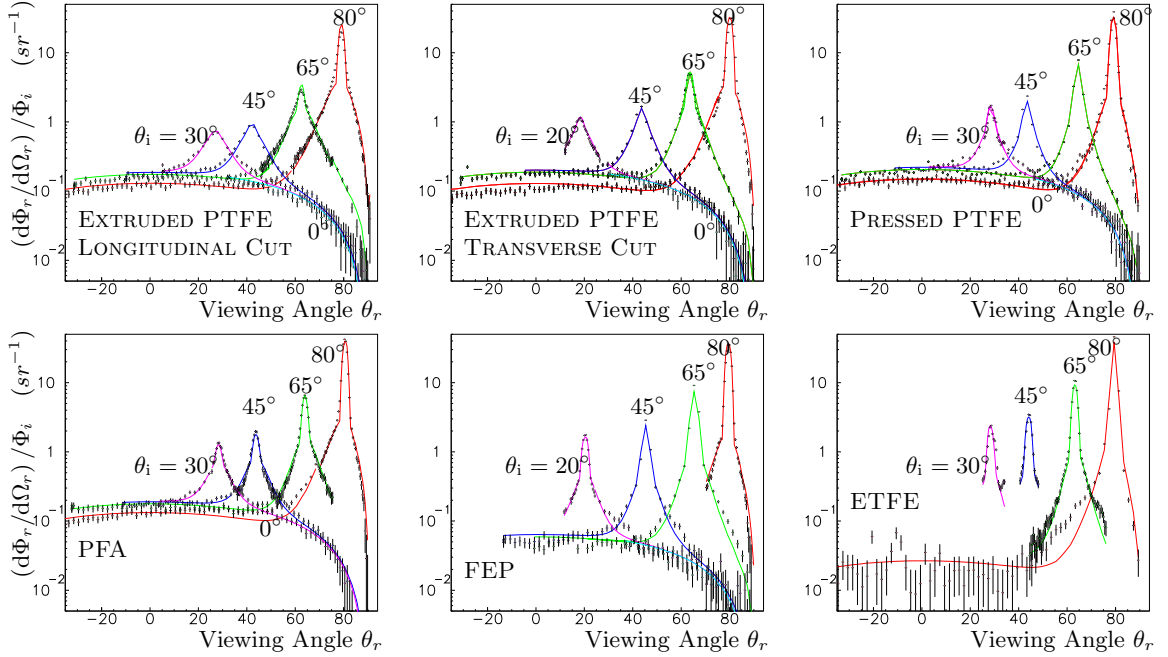}
\caption{Reflectance distributions of various fluoropolymers samples
  for the xenon scintillation light, produced as indicated, plotted as a
function of the viewing angle (in degrees), for various angles of incidence.
Negative values of $\theta_{r}$ mean that $\varphi=\pi$ in figure \ref{sy_model}.
}
\label{manufacture}
\end{figure*}

\begin{table*}[!t]
\vspace{0.5cm}
\begin{center}
\begin{tabular}{lcccc}
\toprule
{\sl Sample}   &$n$    &$\rho_{L}$&$\gamma$ &$K$ \\
               &       &        &             &    \\
\midrule
Extruded ($\perp$) PTFE     & 1.50$\pm$0.03& 0.69$\pm$0.07 & 0.055$\pm$0.007& 3.0$\pm$0.3\\
Extruded ($\parallel$) PTFE & 1.46$\pm$0.04& 0.63$\pm$0.07 & 0.066$\pm$0.008& 4.3$\pm$0.5\\
Pressed PTFE                & 1.45$\pm$0.04& 0.74$\pm$0.07 & 0.049$\pm$0.015& 1.7$\pm$0.2\\
PFA                         & 1.44$\pm$0.04& 0.69$\pm$0.05 & 0.057$\pm$0.006& 2.4$\pm$0.4\\
FEP                         & 1.41$\pm$0.02& 0.22$\pm$0.04 & 0.052$\pm$0.009& 1.2$\pm$0.4\\
ETFE                        & 1.44$\pm$0.03& 0.13$\pm$0.01 & 0.040$\pm$0.010& 1.0$\pm$0.2\\
\bottomrule
\end{tabular}
\caption{Fitted values of $n$, $\rho_L$, $\gamma$ and $K$ for the
  samples measured at 175 nm.  All these surfaces have been polished prior to measurement. The
"Extruded $\perp$" and "Extruded $\parallel$" refer to
cuts perpendicular and parallel to the extrusion direction.}
\label{phemtable}
\end{center}
\end{table*}

We observed the light reflected by samples of fluoropolymers for various incident and viewing 
angles, both in and out of the plane of incidence \cite{silva2}.
Figure \ref{manufacture} shows the reflection
distribution functions for three different samples of polished PTFE and the
copolymers PFA, FEP and ETFE, for the xenon scintillation light (175 nm), in the plane of
incidence. 
The light was produced in a
proportional counter filled with gaseous xenon. The experimental
procedure and the chamber used for the measurements are described
in detail in refs. \cite{silva} and \cite{silva2}. 

The reflectance
distributions of figure \ref{manufacture} show clearly the presence of the three contributions: a diffuse lobe,
a specular lobe and a specular spike. The data was fitted with the function $\varrho$ and the
values of the four unknown parameters extracted, as shown in
table \ref{phemtable}.

\begin{figure}[p] 
  \begin{center}   
    \includegraphics[width=0.85\linewidth]{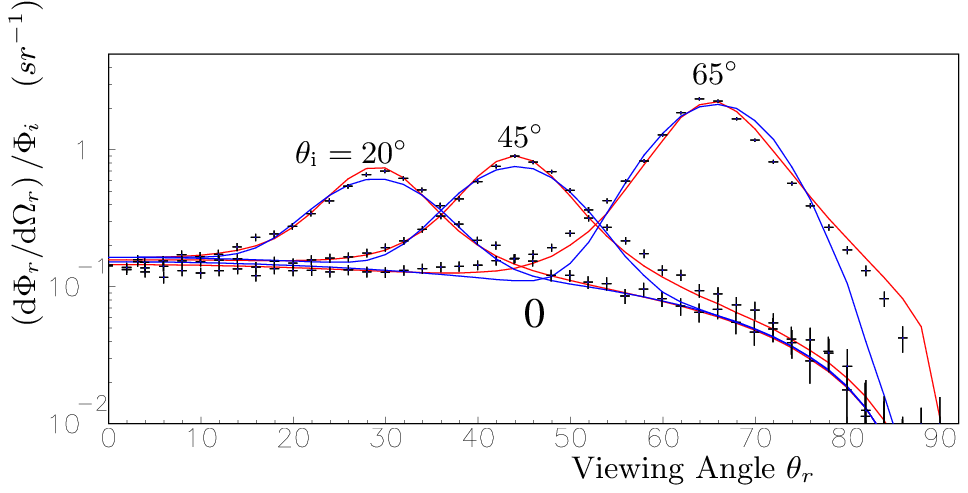}

    \caption{Comparison of the fits to non-polished PTFE using either the 
    Trowbridge-Reitz distribution function (red curves) or the Cook-Torrance
      distribution (blue curves).}
        \label{normal_vstr}
  \end{center}
\end{figure}

In figure \ref{normal_vstr} we compare the fits to a sample of unpolished 
PTFE using either one of the function distributions $P_{TR}$ or $P_{CT}$ 
(eqs. \ref{beck} and \ref{reitz}). The agreement is clearly better if $P_{TR}$ is used.
Therefore, we used this function in all the fits 
shown in the figure \ref{manufacture}.

PTFE and PFA show a pronounced diffuse lobe, corresponding to an albedo between 0.6
to 0.7.
The intensity of this component varies as expected with the angle of
incidence, being suppressed at large angles in result of
decreasing the intensity of the transmitted wave and consequently the sub-surface scattering.

None of the samples measured is a pure diffuser. The coherent spike
increases with the angle as expected and it is especially visible at large angles of
incidence. Its observed width is merely instrumental due to the aperture of the 
incident beam and of the aperture of
the photo-detector used in the experiment.  

The effects of shadowing and masking reduce the
specular lobe, specially at grazing angles, in which 
case the light is scattered towards elsewhere. The geometric factor $G$ referred to above is relevant at
those angles (for angles in excess of $80^\circ$) to reproduce the observed suppression.

\begin{figure}
\centering
\includegraphics[width=2.2in]{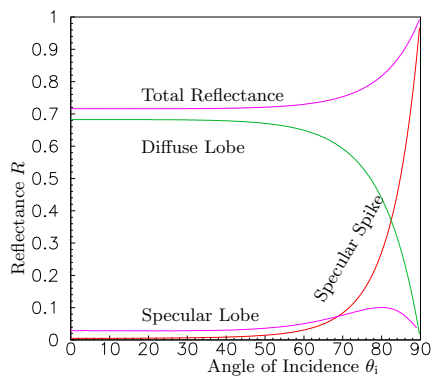}

\caption{The reflectance of pressed polished PTFE as a function of the angle of
  incidence 
${\mathrm{\theta_{i}}}$ for light of $=175$ nm in
  vacuum.}
\label{ptfe_totr}
\end{figure}

The total reflectance of the surface can be calculated for each angle of incidence by
integrating the function $\varrho$ for all viewing directions:
\begin{equation}
R\left(\theta_{i}\right) = \int_{-\pi}^{\pi}
\int_{0}^{\frac{\pi}{2}} \frac{1}{G}
\varrho\left(\theta_{i},\theta_{r},\varphi_{r}\right)
\sin{\theta_{r}} \mathrm{d}\theta_{r}\mathrm{d}\varphi_{r}
\end{equation}
It  is shown in figure \ref{ptfe_totr} for
polished pressed PTFE as a function of the angle of incidence, $\theta_i$.
The three reflection components that are shown remain fairly constant up to 
$\theta_i\sim60^\circ$, the diffuse reflection dominating all along
at 68\%, whereas the specular lobe and specular
spike amount to only 2.8\% and 0.5\% respectively. Above 70$^{\circ}$
the coherent spike is the main component of the specular reflection.

\begin{figure}
\centering 
\includegraphics[width=2.2in]{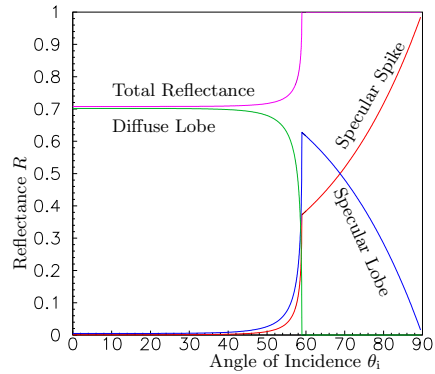}

\caption{The expected reflectance of pressed polished PTFE in contact with
  liquid xenon as a function of the angle of
  incidence ${\mathrm{\theta_{i}}}$ for light of $=175$ nm
  ($n_{\mathrm{Xe}}=1.69 $\cite{solovov}).}
\label{ptfe_totr2}
\end{figure}

This model can be also applied to calculate the reflectance of PTFE in
contact with a liquid instead of vacuum/gas. 
Assuming that the albedo and refraction index of PTFE remain the
same it is straightforward to
conclude that in this case much more light will be reflected by the
surface. The expected reflectance predicted by this model 
 for a PTFE liquid-xenon interface is plotted in figure \ref{ptfe_totr2}.

\section{Conclusion}
The reflectance of polished fluoropolymers exhibit three main reflection
components, a diffuse lobe, a specular lobe and a specular spike.
The process can be modeled with only four free parameters and  
reproduces fairly well the details of the  reflectance distribution measured in these materials.

The data show that the specular lobe is best reproduced if the reflecting surface is
approximated in the model by an ensemble of ellipsoid-shaped
micro-surfaces. This conclusion comes at no surprise though,
  since these materials are granular by nature.

The relative intensity of the coherent specular spike 
appears to vary exponentially with the cosine of the angle of incidence, $\cos\theta_i$. 
This is in contrast to the prediction of the 
Beckmann-Spizzichino model of a dependency in $\cos^2\theta_i$.
This issue deserves additional studies in the future.

In a PTFE vacuum interface the diffuse lobe is dominant for most of
the incident angles. However, the total reflectance computed for PTFE
in contact with liquid xenon shows a much more pronounced contribution
from the specular reflection. The average reflectance is in
agreement with 
previous estimates resulting in values of ~90\%
  for the total reflectance assumed to be independent of the incident
  angle, in this case \cite{yamashi}, \cite{fneves}.

\end{document}